\documentclass[12pt]{article}

\usepackage{amsmath,amsfonts,amssymb,amsthm,mathtools}
\usepackage{graphicx}
\usepackage{url}
\usepackage{multirow}
\usepackage{authblk}
\usepackage[margin=1in]{geometry}
\usepackage{adjustbox}
\usepackage{float}
\usepackage{nicefrac}       
\usepackage{setspace}
\graphicspath{ {./figures/} }
\usepackage{subcaption}
\usepackage{lineno}
\usepackage{comment,caption}
\usepackage{tikz,verbatim}
\usepackage{wrapfig}
\usepackage{lipsum}
\usepackage{mwe}
\usepackage{graphbox}
\usepackage[english]{babel}
\usepackage{csquotes}
\usepackage{array}
\usepackage{bm}
\usepackage{xcolor}
\usepackage{dirtytalk}
\usepackage{threeparttable}
\usepackage{chngcntr}
\usepackage[title]{appendix}
\usepackage{xr-hyper}
\usepackage{hyperref}
\usepackage{natbib}
\makeatletter
\newcommand*{\addFileDependency}[1]{
  \typeout{(#1)}
  \@addtofilelist{#1}
  \IfFileExists{#1}{}{\typeout{No file #1.}}
}
\makeatother
\newcommand*{\myexternaldocument}[1]{%
    \externaldocument{#1}%
    \addFileDependency{#1.tex}%
    \addFileDependency{#1.aux}%
}
\myexternaldocument{supplementary}

\title{A Bayesian joint model for mediation analysis with matrix-valued mediators}

\author[1]{\small Zijin Liu}
\author[1,2]{\small Zhihui (Amy) Liu}
\author[2]{\small Ali Hosni}
\author[2]{\small John Kim}
\author[3]{\small Bei Jiang}
\author[1]{\small Olli Saarela\thanks{Correspondence to: Olli Saarela, Dalla Lana School of Public Health, 155 College Street, Toronto, Ontario M5T 3M7, Canada. Email: \texttt{olli.saarela@utoronto.ca}}}

\affil[1]{\small Dalla Lana School of Public Health, University of Toronto}
\affil[2]{\small Princess Margaret Cancer Centre, University Health Network}
\affil[3]{\small Department of Mathematical and Statistical Sciences, University of Alberta}

\date{\today}

\newcommand{\mvec}{\mathrm{vec}}
\newcommand{\bI}{\bm{I}}

\begin{document}
\maketitle

\begin{abstract}
Unscheduled treatment interruptions may lead to reduced quality of care in radiation therapy (RT). Identifying the RT prescription dose effects on the outcome of treatment interruptions, mediated through doses distributed into different organs-at-risk (OARs), can inform future treatment planning. The radiation exposure to OARs can be summarized by a matrix of dose-volume histograms (DVH) for each patient. Although various methods for high-dimensional mediation analysis have been proposed recently, few studies investigated how matrix-valued data can be treated as mediators. In this paper, we propose a novel Bayesian joint mediation model for high-dimensional matrix-valued mediators. In this joint model, latent features are extracted from the matrix-valued data through an adaptation of probabilistic multilinear principal components analysis (MPCA), retaining the inherent matrix structure. We derive and implement a Gibbs sampling algorithm to jointly estimate all model parameters, and introduce a Varimax rotation method to identify active indicators of mediation among the matrix-valued data. Our simulation study finds that the proposed joint model has higher efficiency in estimating causal decomposition effects compared to an alternative two-step method, and demonstrates that the mediation effects can be identified and visualized in the matrix form. We apply the method to study the effect of prescription dose on treatment interruptions in anal canal cancer patients.

\end{abstract}

\newpage

\section{Introduction}
\label{sec:intro}
In cancer radiation therapy (RT), it is important to avoid unscheduled treatment interruptions because they can compromise tumor control and increase overall treatment time, affecting quality of care \citep{o2022compensation,hendry1996modelled}. Understanding the causes of treatment interruptions can assist radiation oncologists in better planning RT strategies. The primary goal of RT planning is to deliver the prescription dose to the target tumor while minimizing damage to surrounding healthy tissues, also called organs-at-risk (OARs). Excessive radiation exposure to OARs can cause toxicity related complications, such as skin burns, hair loss, and digestive problems, which can lead to these unscheduled treatment interruptions. Identifying the dose effects on the outcome of treatment interruptions that are mediated through OARs, and understanding how these effects are distributed into different OARs and their dose regions, can inform future treatment planning. This suggests a rationale for causal mediation analysis, with the RT prescription dose as the treatment, the radiation exposure to OARs as the mediator, and the treatment interruption as the outcome. We are particularly interested in which OARs and doses mediate this relationship.

The radiation exposure to a specific OAR can be represented at voxel level as a function of three-dimensional spatial location. These voxel specific doses can be summarized by a distribution function, the discretized version of which is known as the dose-volume histogram (DVH), which relates the volume of an OAR and doses. DVH data are high-dimensional because the dose range can be discretized into a long vector ($q$-dimensional) of small intervals with each cut-off value as a variable. DVH data are also highly correlated because voxels that are spatially close receive similar doses, resulting in a spatially correlated dose distribution. The DVH data vectors for all $p$ OARs of interest can be combined into matrix-valued data of $p \times q$ matrix per patient. In this matrix, each column represents a dose cut-off value, each row corresponds to a specific OAR, and each entry is the volume of the specific OAR receiving at least the particular dose. Several authors have used principal component analysis (PCA) to extract features from DVH data to predict complications \citep{dawson2005use,bauer2006principal,skala2007patient,dean2016functional}. However, they typically analyze only one OAR/DVH at a time.

Conventional mediation analysis involves a single or at most a few mediators. Structural Equation Models (SEM) are commonly used, incorporating a mediator model that regresses the mediator on the treatment and an outcome model that regresses the outcome on both the treatment and the mediator. When the number of mediators exceeds the sample size, the problem is considered high-dimensional, and mediation analysis methods for high-dimensional and correlated mediators have been developed in recent years. \citet{huang2016hypothesis} adapted the dimension reduction idea from PCA. They proposed to transform the original high-dimensional mediators into a new set of uncorrelated, lower-dimensional mediators, which subsequently replace the original mediators in the SEM. \citet{zhao2020sparse} improved this work by using sparse PCA to obtain more interpretable transformed mediators. \citet{zhang2016estimating} proposed a penalized regression for the outcome model to select the mediators that have large effects on the outcome, and then fit the mediator model only on those selected mediators. Similar methods can be found in \citet{gao2019testing} and \citet{zhang2021high} with different penalty functions. 

All of the above methods involve a two-step process, where the outcome model and the mediator model are fitted separately. Alternatively, \citet{chen2018high} proposed a likelihood-based method to transform the original mediators into a small number of uncorrelated mediators, and then jointly estimated all model parameters, including transformation loadings. \citet{song2020bayesian, song2021bayesian} proposed a Bayesian sparse linear mixed model to jointly estimate mediation and outcome model parameters. With continuous shrinkage priors to place penalty, they achieved sparse estimated mediation effects. \citet{derkach2019high} proposed a latent variable joint model that assumes the treatment directly influences a few conditionally independent latent factors acting as mediators, and aimed for sparsity through a penalty. Readers are referred to \citet{clark2023methods} for a more comprehensive review of high-dimensional mediation analysis methods.

The naive way to analyze the $p \times q$ matrix-valued data is to vectorize it into a $pq \times 1$ vector, and then apply conventional dimension reduction techniques, such as PCA. However, this ignores the row and column structure of the matrix, causing the loss of the inherent matrix structure information. A large body of literature investigated how to handle matrix-valued data as covariates in regression settings \citep{zhou2013tensor, zhao2014structured, zhou2014regularized, hoff2015multilinear, li2017parsimonious, ding2018matrix}, and these methods could potentially be adapted to mediation analysis. For dimension reduction, Multilinear Principal Component Analysis (MPCA), an extension of PCA for matrix-valued or even tensor-valued data, has been proposed by \citet{lu2008mpca}, with its statistical properties further studied by \citet{hung2012multilinear}. However, these methods are primarily algorithm-based, and integrating them into the causal mediation framework would require a two-step approach of first extracting low-dimensional features using MPCA, and using these features as new mediators to fit SEM.

\citet{tipping1999probabilistic} proposed a probabilistic PCA (PPCA), which shares a formulation similar to factor analysis, but constrains the residual variances to be equal across all variables. \citet{ding2014dimension} further extended PPCA to a dimension-folding PCA, which essentially is a probabilistic MPCA for matrix-valued data. \citet{jiang2020bayesian} proposed a Bayesian joint prediction model, where they incorporated probabilistic MPCA as a submodel and used the features from MPCA to formulate a probit model as another submodel. This motivates us to consider incorporating MPCA into the mediation analysis framework using a joint model structure similar to \citet{derkach2019high}.

In this paper, we propose a novel Bayesian joint mediation model for matrix-valued DVH data, which is constituted by three components: (i) probabilistic MPCA for matrix-valued data, (ii) mediator model for the features from MPCA, and (iii) a probit model for a binary outcome. We derive a Monte Carlo Markov Chain (MCMC) algorithm to jointly estimate all model parameters, and introduce a simple rotation method to identify active indicators of mediation among the matrix-valued DVH data. Although our joint model is motivated by DVH data, the proposed methods can be applied to any matrix-valued mediator problem. 

The rest of this paper are structured as follows: In section 2, we formulate the proposed joint mediation model, introduce a quantity measuring mediation, and define the causal decomposition effects. We then assess the identifiability of the model, outline an MCMC algorithm for model estimation, and also introduce an alternative two-step estimation method. In section 3, we conduct simulation studies to assess the performance of causal decomposition estimates, as well as the ability of the joint model to identify active indicators of mediation. In section 4, we apply our proposed joint model to a DVH dataset from a prospective cohort study, and then suggest a method to select the number of latent features in the proposed joint model. Section 5 concludes our findings with a discussion. 

\section{Model}
\label{s:model}
\subsection{Joint Mediation Model}

For each patient $i=1,\cdots,n$, we denote $\bm{X}_i \in \mathbb{R}^{p \times q}$ as the $p \times q$ dimensional matrix-valued data, $E_i$ as the exposure/treatment, $\bm{Z}_i=(Z_{i1},\cdots,Z_{iK})^\top  \in \mathbb{R}^{K \times 1}$ as a vector of covariates, and $Y_i$ as a binary outcome. Since $\bm{X}_i$ is generally high-dimensional, we aim to find a latent, lower-dimensional $p_0 \times q_0$ matrix $\bm{T}_i \in \mathbb{R}^{p_0 \times q_0} (p_0 \leq p, q_0 \leq q)$ to represent $\bm{X}_i$, while still preserving its matrix structure. Motivated by MPCA, we formulate the first component of our joint model as
\begin{equation}
    \bm{X}_i= \bm{\mu} + \bm{A} \bm{T}_i \bm{B}^\top + \bm{\varepsilon}_{X,i} \label{equ:JointMedMPCA}
\end{equation}
Latent features $\bm{T}_i \in \mathbb{R}^{p_0 \times q_0}$ can be seen as mapping $\bm{X}_i$ through a two-way projection, where $\bm{A} \in \mathbb{R}^{p \times p_0} $ is a row-wise orthonormal projection matrix such that $\bm{A}^\top \bm{A} = \bI_{p_0}$ and $\bm{B}\in \mathbb{R}^{q \times q_0}$ is a column-wise orthonormal projection matrix such that $\bm{B}^\top \bm{B} = \bI_{q_0}$. The error matrix $\bm{\varepsilon}_{X,i} \in \mathbb{R}^{p \times q}$ is assumed to be element-wise independent and independent of $\bm{T}_i$. Letting $\text{vec}(\cdot)$ denote the vectorization operator, we assume $\mvec(\bm{\varepsilon}_{X,i}) \sim N(\bm{0},\phi^{-1} \bI_{pq})$, where $\phi$ is a precision parameter. Finally, we define $\bm{\mu} \in \mathbb{R}^{p \times q}$ as the mean matrix of $\bm{X}_i$. 

\textit{Remark 1:} Model \eqref{equ:JointMedMPCA} is equivalent to $\mvec (\bm{X}_i) = \mvec(\bm{\mu}) + (\bm{B} \otimes \bm{A}) \mvec(\bm{T}_i) + \mvec (\bm{\varepsilon}_{X,i})$, by the property of the Kronecker product denoted by $\otimes$. If we let $\bm{W}=\bm{B} \otimes \bm{A}\in \mathbb{R}^{pq \times p_0q_0}$, we can see that MPCA constrains the parameter space of the loading matrix for $\mvec (\bm{X}_i)$ by a Kronecker envelope with $pp_0+qq_0$ free parameters. Compared to a PPCA \citep{tipping1999probabilistic} formulation $\mvec (\bm{X}_i) = \mvec(\bm{\mu}) + \bm{\Lambda} \mvec(\bm{T}_i) + \mvec (\bm{\varepsilon}_{X,i})$, where $\bm{\Lambda}$ is unconstrained with $pq \times p_0q_0$ parameters, MPCA greatly reduces the number of parameters.

The second component of our joint model is to relate latent features $\mvec(\bm{T}_i)$ to the treatment $E_i$ and covariates $\bm{Z}_i$. Let $\bm{\beta}_{ET} = (\beta_{ET,1},\cdots,\beta_{ET,p_0q_0})^\top$, and $\bm{\varepsilon}_{T,i}$ be a $p_0 \times q_0$ error matrix. The mediator model is specified as
\begin{align}
& \mvec (\bm{T}_i) = \bm{\beta}_{ET}E_i + \bm{\Omega}_{ZT} \bm{Z}_i  + \mvec(\bm{\varepsilon}_{T,i}) \label{equ:JointMedT}
\end{align}
where $\bm{\Omega}_{ZT}=(\bm{\varOmega}_1,\cdots,\bm{\varOmega}_K)\in \mathbb{R}^{p_0q_0 \times K}$ is a matrix of parameters for $k=1,\cdots,K$ covariates, with $\bm{\varOmega}_k = (\Omega_{1,k},\cdots,\Omega_{p_0q_0,k})^\top$ for a total of $p_0q_0$ features, and $\mvec(\bm{\varepsilon}_{T,i}) \sim N (\bm{0},\bI_{p_0q_0})$.

Finally, the last component of the joint model is to link the binary outcome by a probit model. Let $\bm{\beta}_{TY} = (\beta_{TY,1},\cdots,\beta_{TY,p_0q_0})^\top$ and $\bm{\beta}_{ZY} =(\beta_{ZY,1},\cdots,\beta_{ZY,K})^\top$, then
\begin{equation} \label{equ:JointMedProbit}
\Phi^{-1}[\text{Pr}(Y_i=1 \mid E_i, \bm{T}_i, \bm{Z}_i)]=\alpha_Y+\beta_{EY}E_i+\bm{\beta}_{TY}^\top \mvec(\bm{T}_i) + \bm{\beta}_{ZY}^\top \bm{Z}_i
\end{equation}
where $\Phi(\cdot)$ represents the cumulative distribution function of $N(0,1)$. 

\begin{figure}
    \centerline{
    \includegraphics[width=0.7\textwidth]{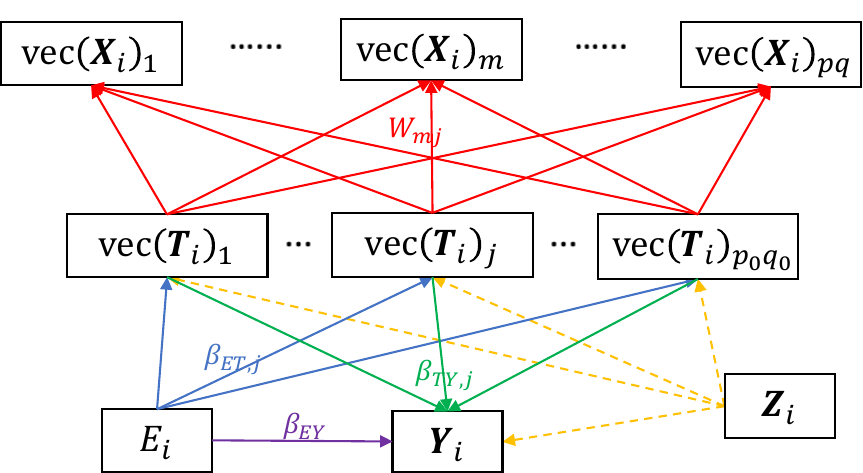}}
    \caption{DAG of the proposed joint model. $E_i$ is treatment, in the motivating example the radiotherapy prescription dose. $Y_i$ is binary outcome, unplanned treatment interruption. For $m=1,\cdots,pq$, $\mathrm{vec} (\bm{X}_i)_m$ is the $m$th element of the vectorized matrix-valued indicators of mediation for individual $i$, reflecting a dose bin of the DVH for a specific OAR. For $j=1,\cdots,p_0q_0$, $\mathrm{vec} (\bm{T}_i)_j$ is the $j$th latent feature. $W_{mj}$ (red) is loading of $\mathrm{vec} (\bm{X}_i)_m$ on $\mathrm{vec} (\bm{T}_i)_j$. $\beta_{ET,j}$ (blue) is the regression coefficient between the treatment and latent feature $j$, $\beta_{TY,j}$ (green) regression coefficient between the latent feature $j$ and outcome, and $\beta_{EY}$ (purple) regression coefficient between the treatment and outcome. Dashed yellow arrows reflect covariate effects.}
    \label{fig:DAGjointMed}
\end{figure}

The proposed joint model (\ref{equ:JointMedMPCA}-\ref{equ:JointMedProbit}) suggest a causal relationship between the treatment $E_i$, latent features $\bm{T}_i$, covariates $\bm{Z}_i$, matrix-valued data $\bm{X}_i$, and the outcome $Y_i$. This relationship can be described by the directed acyclic graph (DAG) in Figure \ref{fig:DAGjointMed}. Here the latent features are conceptualized as the mediators, while $\bm{X}_i$ has the role of a manifest/indicator variable, with arrows/edges pointing into it. 

We now introduce a quantity to identify which elements in the matrix-valued data $\bm{X}_i$ are active indicators of mediation in the treatment-outcome relationship. Let $W_{mj}$ be the $(m,j)$th element of $\bm{W}=\bm{B} \otimes \bm{A}$, where $m=1,\cdots,pq$ and $j=1,\cdots,p_0q_0$. We define the active indicators among the total of $pq$ elements in $\bm{X}_i$ as the set of $\{m: \sum_j |\beta_{ET,j} \beta_{TY,j} W_{mj}| \neq 0 \}$, similar to \citet{derkach2019high}. To motivate this, active indicators in $\bm{X}_i$ should be related to both $E_i$ and $Y_i$ through at least one of the $p_0q_0$ latent features. This gives a non-zero value in at least one of the products $|\beta_{ET,j}\beta_{TY,j}W_{mj}|$ for $j=1,\cdots,p_0q_0$. In practice, we can select the set $\{m: \sum_j |\widehat{\beta}_{ET,j} \widehat{\beta}_{TY,j} \widehat{W}_{mj}| \neq 0 \}$ as the active indicators of mediation. The quantity 
\begin{equation}\label{eq:med}
\sum_{j=1}^{p_0q_0} |\beta_{ET,j} \beta_{TY,j} W_{mj}|, \qquad m=1,\cdots,pq, 
\end{equation}
can be mapped back to the original $p \times q$ matrix structure to obtain interpretations for matrix-valued mediators. Posterior probabilities $\text{Pr}(\sum_j |\beta_{ET,j} \beta_{TY,j} W_{mj}| > \kappa \mid \mathcal D) $, where $\kappa$ is a chosen threshold and $\mathcal D$ represents all observed data, can also be used and reshaped to the original matrix form for a probabilistic interpretation. Our model can be seen as extension of \citet{derkach2019high} to matrix-valued mediators, but the added advantage of the Bayesian estimation procedure is that we can obtain probabilistic inferences for all the quantities of interest, including the above defined measures for the strength of mediation.

While we envision that the results of the high-dimensional mediation analysis will primarily be interpreted through quantities \eqref{eq:med}, the joint model can be used to estimate causal mediation analysis decomposition of the total effect (TE) into natural indirect (NIE) and direct (NDE) effects; details are given in Web Appendix \ref{appdx:Der_Effects_expect}.

\subsection{Identifiability}
Let $\Theta = (\bm{A},\bm{B},\bm{\mu},\bm{\beta}_{ET},\bm{\Omega}_{ZT},\alpha_Y,\beta_{EY},\bm{\beta}_{TY},\bm{\beta}_{ZY},\phi)$ denote all parameters in our joint model (\ref{equ:JointMedMPCA}-\ref{equ:JointMedProbit}). We can show that $\Theta$ is identifiable up to, but not including, an orthogonal rotation. That is, for any orthonormal matrices $\bm{P} \in \mathbb{R}^{p_0 \times p_0}$ such that $\bm{P}^\top\bm{P} = \bm{P}\bm{P}^\top = \bI_{p_0}$ and $\bm{Q} \in \mathbb{R}^{q_0 \times q_0}$ such that $\bm{Q}^\top\bm{Q} = \bm{Q}\bm{Q}^\top = \bI_{q_0}$, we let $\widetilde{\bm{A}} = \bm{A} \bm{P}$, $\widetilde{\bm{B}} = \bm{B} \bm{Q}$, $\widetilde{\bm{\beta}}_{ET} = (\bm{Q} \otimes \bm{P})^\top \bm{\beta}_{ET} $, $\widetilde{\bm{\beta}}_{TY} = (\bm{Q} \otimes \bm{P})^\top \bm{\beta}_{TY}$, $\widetilde{\bm{\Omega}}_{ZT} = (\bm{Q} \otimes \bm{P})^\top \bm{\Omega}_{ZT}$, and $\mvec (\widetilde{\bm{T}}_i) = (\bm{Q} \otimes \bm{P})^\top \mvec (\bm{T}_i)$, the full likelihood of the joint model (see Web Appendix \ref{appdx:likelihood}) will be the same for the two sets of parameters $\Theta$ and $\widetilde{\Theta}=(\widetilde{\bm{A}},\widetilde{\bm{B}},\bm{\mu},\widetilde{\bm{\beta}}_{ET},\widetilde{\bm{\Omega}}_{ZT},\alpha_Y,\beta_{EY},\widetilde{\bm{\beta}}_{TY},\bm{\beta}_{ZY},\phi)$. Identifiability up to the orthogonal rotation follows because the column spaces (spans) of the loading matrices $\bm{A}$ and $\bm{B}$ are identifiable (\citealp{hung2012multilinear, ding2014dimension}), and the orthonormal constraint further restricts the rotations to orthogonal ones. Details are given in Web Appendix \ref{appdx:identify_params}.  

Free orthogonal rotations of model parameters causes the quantity \eqref{eq:med}
to also be identifiable only up to orthogonal rotations. To solve this issue and identify the active indicators of mediation, \citet{derkach2019high} used L1 penalization to shrink $\bm{\beta}_{ET}$, $\bm{\beta}_{TY}$ and $\bm{W}$ to exact zero for those true zero parameters, such that the estimated mediation quantities $\sum_j |\widehat{\beta}_{ET,j} \widehat{\beta}_{TY,j} \widehat{W}_{mj}|$ can also be penalized to zero for those inactive indicators (i.e. $\{ m: \sum_j |\beta_{ET,j} \beta_{TY,j} W_{mj}| = 0 \}$). Instead, we propose a much simpler remedy by fixing the orthogonal rotation using the Varimax principle, which is described in detail in Section \ref{sec:gibbs}.

We note that bi-products of any two combination of $\bm{W}$,  $\bm{\beta}_{ET}$, $\bm{\Omega}_{ZT}$, and $\bm{\beta}_{TY}$, such as $\bm{W}\bm{W}^\top$, $\bm{\beta}_{TY}^\top \bm{\beta}_{TY}$, $\bm{W}\bm{\beta}_{ET}$, $\bm{\beta}_{ET}^\top \bm{\beta}_{TY}$, and $\bm{\beta}_{TY}^\top \bm{\Omega}_{ZT}$, are identifiable and not influenced by orthogonal rotations. Furthermore, $\beta_{EY}$ and $\bm{\beta}_{ZY}$ are uniquely identifiable, and Web Appendix \ref{appdx:identy_CDE} shows that the overall NIE, NDE and TE are all uniquely identifiable. Individual indirect pathways NIE$_1$ to NIE$_{p_0q_0}$ are only identifiable up to orthogonal rotations, because indirect individual pathway bi-products $\beta_{ET,j}\beta_{TY,j}$ for $j=1,\cdots,p_0q_0$ are only identifiable up to orthogonal rotations even under the ignorability assumption \eqref{equ:Ignorability4} provided in Web Appendix \ref{appdx:Der_Effects_expect}. This is not a serious limitation as we do not aim to attach interpretation to the individual latent variables (rather, interpretation is sought through the mediation quantities \eqref{eq:med}).

\subsection{Gibbs Sampling and Varimax Rotation}
\label{sec:gibbs}

Building on a Bayesian estimation approach proposed by \citet{jiang2020bayesian} for matrix-valued predictors, we propose a Gibbs sampling algorithm to estimate parameters in our joint mediation model. Similar to \citet{jiang2020bayesian}'s algorithm, the dimension of the MPCA model is further restricted to $p<p_0,q<q_0$ (instead of $p\leq p_0,q\leq q_0$ in submodel \eqref{equ:JointMedMPCA}). That is, for computational reasons we reduce both dimensions of the matrix-valued data. This restriction is discussed in Section \ref{sec:Discussion}. The full conditional posterior distributions for model parameters, and the details of the Gibbs sampling algorithm can be found in Web Appendix \ref{appdx:MCMC}. 

Because both $\bm{A}$ and $\bm{B}$ are orthonormal matrices, we naturally select the uniform distribution in the space of orthonormal matrices (i.e., Stiefel manifolds) as prior distributions for $\bm{A}$ and $\bm{B}$. If a random matrix follows a uniform distribution on Stiefel manifolds, then each column of this random matrix can be represented by a conditional distribution given other columns \citep{hoff2007model}. By applying this property, the posterior samples of $\bm{A}$ and $\bm{B}$ can be drawn column-by-column based on their full conditional posterior distributions given on all other parameters and other columns of themselves. For other model parameters, we select weakly informative conjugate priors to facilitate the Gibbs sampling and to impose a mild penalty on large regression parameter estimates to obtain more stable results. 

To identify active indicators of mediation, we propose to modify the original Gibbs sampling algorithm by fixing the orthogonal rotation using the Varimax principle. Specifically, after each iteration of Gibbs sampling, we (1) implement Varimax rotation for $\bm{A}$ and $\bm{B}$ after the last step of the original Gibbs sampling algorithm described in Web Appendix \ref{appdx:MCMC}, and record the rotation matrices $\bm{P}$ and $\bm{Q}$; (2) calculate the rotated $\widetilde{\Theta}$ and $\mvec (\widetilde{\bm{T}}_i)$ before the next iteration of Gibbs sampling as $\widetilde{\bm{A}} = \bm{A} \bm{P}$, $\widetilde{\bm{B}} = \bm{B} \bm{Q}$, $\widetilde{\bm{\beta}}_{ET} = (\bm{Q} \otimes \bm{P})^\top \bm{\beta}_{ET} $, $\widetilde{\bm{\beta}}_{TY} = (\bm{Q} \otimes \bm{P})^\top \bm{\beta}_{TY}$,$\widetilde{\bm{\Omega}}_{ZT} = (\bm{Q} \otimes \bm{P})^\top \bm{\Omega}_{ZT}$, and $\mvec (\widetilde{\bm{T}}_i) = (\bm{Q} \otimes \bm{P})^\top \mvec (\bm{T}_i)$, keeping other parameters unchanged; and (3) use $\widetilde{\Theta}$ and $\mvec (\widetilde{\bm{T}}_i)$ instead of $\Theta$ and $\mvec (\bm{T}_i)$ to be sent into the next iteration of Gibbs sampling, until reach the maximum iteration rounds.

The motivation is, for each column of $\bm{W}$, we only want to keep a small number of elements with large values, while forcing most of them close to zero. This makes $\bm{W}$ sparse, which helps the joint model identify active indicators of mediation, rather than capturing arbitrary noise produced by random orthogonal rotations. \citet{rohe2020vintage} recently showed Varimax rotated PCA loadings are identifiable up to a column permutation and sign flip under some mild assumptions (Theorem 6.1 in their paper). This gives identifiable mediation quantities because \eqref{eq:med} is invariant after column reordering and sign changes of $\bm{W}$. In our simulation study, we demonstrate that \eqref{eq:med} is identifiable if the data generating mechanism follows the model with Varimax rotated loadings. An alternative two-step estimation method is described in Web Appendix \ref{sec:two_step}.

\section{Simulation Study}
\subsection{Setup}
\label{sec:simulation_step}

A simulation study was conducted to evaluate our proposed model performance in two aspects: (i) To compare the performance of causal decomposition estimates of NIE, NDE, and TE from the joint model to the estimates from the two-step method, and (ii) To evaluate the capability of the joint model to identify active indicators of mediation. Additional simulation studies are reported in Web Appendix \ref{appdx:compare_study} and \ref{appdx:misStudy}, where we compare the MPCA-based models to PCA-based models for vectorized data, as well as study the effects of overspecifying the dimension of the dimension reduction.

For the data generating mechanism, we consider a low dimensional matrix $(p,q)=(10,10)$ and a high-dimensional matrix $(p,q)=(10,50)$. In both scenarios, we fix $(p_0,q_0)=(2,2)$, which assumes the original $10 \times 10$ and $10 \times 50$ matrix-valued data can be fully represented by a $2 \times 2$ matrix of features. We then let $\bm{\beta}_{TY}=(0.7,0,0.7,0)$, $\bm{\beta}_{ET}=(0.5,0.5,0,0)$, which assumes only the first vectorized latent feature is associated with both $E_i$ and $Y_i$, while the second is only associated with $E_i$, the third is only associated with $Y_i$, and the last does not relate to either. $\bm{A}$ and $\bm{B}$ are first (i) uniformly generated from the space of orthonormal matrices (Stiefel manifolds). To create inactive indicators while keeping some active indicators of mediation, we need to impose sparseness into $\bm{A}$ and $\bm{B}$. We (ii) apply Varimax rotations to both generated $\bm{A}$ and $\bm{B}$, and only keep elements that are greater than 0.3, while assign 0 to elements that are less than or equal to 0.3. Finally, (iii) for each column of $\bm{A}$ and $\bm{B}$, we normalize it to a unit vector by dividing the square root of the dot product of it. We note that the steps (i)-(iii) does not necessarily generate orthonormal $\bm{A}$ and $\bm{B}$, because small correlations may exist between columns after assigning 0 to them, but otherwise they are similar to Varimax rotated orthonormal matrices with sparseness. For simplicity, we do not consider any covariates. Other model parameters are fixed at $\mvec (\bm{\mu}) =\bm{0}$, $\alpha_Y=0$, $\beta_{EY}=0.3$, and $\phi=25$. For Bayesian hyperparameters, we assume $\phi^{-1} \sim \text{Gamma}(0.1,0.1)$, $\alpha_Y \sim \beta_{EY} \sim N(0,1)$, and $\bm{\beta}_{ET} \sim \bm{\beta}_{TY} \sim N(\bm{0},\bI_{p_0q_0})$.

For two sample size settings $n \in \{100, 300\}$, we simulate 500 replicates of data $\{(Y_i,E_i,\bm{T}_i, \allowbreak \bm{X}_i): i=1,\dots,n\}$. Data are generated as follows given the fixed model parameters $\Theta$: (1) generate $E_i \sim \text{Bernoulli}(0.5)$, and then generate $\mvec(\bm{T}_i) \sim N(\bm{\beta}_{ET}E_i,\bI_{p_0q_0})$; and (2) generate $\mvec(\bm{X}_i) \sim N( \mvec (\bm{\mu}) + (\bm{B} \otimes \bm{A}) \mvec(\bm{T}_i), \phi^{-1}\bI_{pq}) $, and $Y_i \sim \text{Bernoulli}(\Phi[\alpha_Y+\beta_{EY}E_i+\bm{\beta}_{TY}^\top \mvec(\bm{T}_i)])$.

For each simulated data replicate, we apply our Bayesian joint model and the two-step method to obtain the estimates of NIE, NDE and TE. For the Bayesian method, posterior samples for all model parameters are obtained by retaining every 5th draw from 10,000 iterations of MCMC, after a burn-in period of 3000 iterations, resulting in a total of $R=2000$ posterior samples. For every set of posterior sample parameters from $r=1,\dots,R$, we draw $g=1,\cdots,p_0q_0+1$ sets of mediator values $\bm{t} \in \mathbb{R}^{p_0q_0 \times 1}$ for Monte Carlo integration with sample size $S=5000$. Specifically, the $g$th set of $\bm{t}$ is simulated from $N(\bm{\beta}_{ET} \circ \bm{e}_g,\bm{I}_{p_0q_0})$, and $\bm{e}_g$ is the $g$th element of $\{ (0,0,\cdots,0),\allowbreak (1,0,\cdots,0), (1,1,\cdots,0),\cdots,(1,1,\cdots,1) \}$. These $p_0q_0+1$ sets correspond to the different levels of intervention on the mediators in order to calculate the decomposition effects $\textrm{NIE}_j$ and $\textrm{NDE}$, i.e. intervening on each one of the $p_0q_0$ in turn and the non-intervened case. Then, we substitute $\widehat{\alpha}^{(r)}_Y$, $\widehat{\beta}^{(r)}_{EY}$, $\widehat{\bm{\beta}}^{(r)}_{ET}$, $\widehat{\bm{\beta}}^{(r)}_{TY}$, and samples of $\bm{t}^{(s)}$ into the Monte Carlo estimate (Equation \eqref{equ:MC_est} in Web Appendix \ref{appdx:Der_Effects_expect}) to calculate $\text{NIE}^{(r)}$, $\text{NDE}^{(r)}$, and $\text{TE}^{(r)}$. Finally, we average on these $R$ posterior samples to obtain posterior mean NIE, NDE, and TE. For the two-step method, after obtaining estimates of $\widehat{\alpha}_Y$, $\widehat{\beta}_{EY}$, $\widehat{\bm{\beta}}_{ET}$, $\widehat{\bm{\beta}}_{TY}$, we calculate the decomposition effects similarly by Monte Carlo integration. Mean squared errors (MSE), variances, and biases of causal decomposition estimates are calculated to compare these two methods in both $(p,q)=(10,10)$ and $(10,50)$ scenarios.

To identify active indicators of mediation, for the Bayesian joint model, the estimated quantities $\sum_{j=1}^{p_0q_0} |\widehat{\beta}^{(r)}_{ET,j} \widehat{\beta}^{(r)}_{TY,j} \widehat{W}^{(r)}_{mj}|$, $m=1,\cdots,pq$ are calculated for each MCMC iteration after the Varimax rotation, and their posterior means are obtained by averaging over all $R$ samples. Posterior probabilities for all elements of the matrix are calculated by the empirical proportions $\sum_{r=1}^{R} \mathcal{I}[ \allowbreak \sum_{j=1}^{p_0q_0} |\widehat{\beta}^{(r)}_{ET,j} \widehat{\beta}^{(r)}_{TY,j} \widehat{W}^{(r)}_{mj}| > \kappa]  / R$, $m=1,\cdots,pq$, where $\mathcal{I}[\cdot]$ is the indicator function, and $\kappa \in \{ 0.05, 0.1, 0.15 \}$ for $(p,q)=(10,10)$ and $\kappa \in \{ 0.03, 0.06, 0.09 \}$ for $(p,q)=(10,50)$. We note that the posterior probabilities are essentially the sensitivities for active indicators, and one minus specificity for inactive indicators, and thus AUC values could be calculated by varying all possible thresholds $\kappa$. For the two-step method, quantities $\sum_{j=1}^{p_0q_0} |\widehat{\beta}_{ET,j} \widehat{\beta}_{TY,j} \widehat{W}_{mj}|$, $m=1,\cdots,pq$ are calculated using the point estimates of the model parameters, and results are reported in the Web Appendix \ref{appdx:comparativeStudy_results}, Web Figure \ref{fig:appdx_PCA_scatter}. Reported mediation quantities, posterior probabilities, and AUC values are all averaged over 500 simulation replicates.

\subsection{Results}

\begin{table}[ht]
\centering
\caption{\label{tb:CausalEffects} Comparison between different models} 
\begin{threeparttable}
\centering
\begin{tabular}{cc|cccc|cccc}
\hline \hline 
$\bm{n}$ & \textbf{Effects} & \textbf{MSE}\tnote{1} & \textbf{Var}\tnote{1} & \textbf{Bias}\tnote{1} & \textbf{Estimate}\tnote{2} & \textbf{MSE}\tnote{1} & \textbf{Var}\tnote{1} & \textbf{Bias}\tnote{1} & \textbf{Estimate}\tnote{2} \\
\hline
& &\multicolumn{4}{c|}{\textbf{Joint Model: $(p,q)=(10,10)$}}  &\multicolumn{4}{c}{\textbf{Two-step Method: $(p,q)=(10,10)$}} \\
\hline
\multirow{3}{*}{100}  
& NIE & 1.878 & 1.861 & -4.512 & 0.0886 & 3.415 & 3.416 & -2.319 & 0.0908 \\ 
& NDE & 5.846 & 5.842 & -3.997 & 0.0804 & 7.900 & 7.856 &  7.689 & 0.0921 \\         
& TE  & 6.337 & 6.277 & -8.509 & 0.1690 & 9.175 & 9.164 &  5.371 & 0.1829 \\    
\hline
\multirow{3}{*}{300} 
& NIE & 0.643 & 0.644 & -0.882 & 0.0922 & 1.125 & 1.119 & -2.856 & 0.0902 \\ 
& NDE & 1.929 & 1.930 & -1.763 & 0.0826 & 2.161 & 2.139 &  5.139 & 0.0895 \\   
& TE  & 1.958 & 1.954 & -2.645 & 0.1749 & 2.523 & 2.523 &  2.284 & 0.1798 \\    
\hline
& &\multicolumn{4}{c|}{\textbf{Joint Model: $(p,q)=(10,50)$}}  &\multicolumn{4}{c}{\textbf{Two-step Method: $(p,q)=(10,50)$}} \\
\hline
\multirow{3}{*}{100}  
& NIE & 2.309 & 2.216 & -10.223 & 0.0829 & 3.671 & 3.656 & -5.746 & 0.0874 \\ 
& NDE & 6.235 & 5.998 & -16.323 & 0.0681 & 7.959 & 7.951 & -6.958 & 0.0774 \\ 
& TE & 6.728 & 6.054 & -26.546 & 0.1510 & 8.265 & 8.144 & -12.705 & 0.1648 \\ 
\hline
\multirow{3}{*}{300} 
& NIE & 0.574 & 0.574 & -1.874 & 0.0912 & 1.013 & 0.978 & -6.352 & 0.0867 \\ 
& NDE & 2.603 & 2.615 & 1.121 & 0.0855 & 2.925 & 2.872 & 8.198 & 0.0926 \\ 
& TE & 2.462 & 2.474 & -0.753 & 0.1768 & 3.454 & 3.468 & 1.845 & 0.1793 \\ 
\hline \hline
\end{tabular}
\begin{tablenotes}
\small
\item[1] The values displayed for MSE, bias, and variance are the actual values $\times$ 1000; MSE = Mean Squared Error; Var = Variance
\item[2] True NIE = 0.0931; True NDE = 0.0844; True TE = 0.1775
\end{tablenotes}
\end{threeparttable}
\end{table}

Table \ref{tb:CausalEffects} summarizes the performance (evaluated by MSEs, variances and biases) of NIE, NDE and TE estimators over the 500 simulation replicates under two sample size settings and low and high dimensional scenarios for the joint model and two-step method. We observe that for all scenarios across all decomposition effects, the joint model provides lower MSEs and variances compared to the two-step method. This is expected because the two-step method carries errors from the first step estimating MPCA loadings to the second step estimating regression parameters, leading to additional inaccuracies. In the joint model, all model parameters are instead estimated simultaneously through a joint likelihood, which allows information sharing and thus reduces estimation errors. Increasing the sample size $n$ from 100 to 300 reduces the MSEs, variances, and biases for almost all estimates, except the biases of NIE and NDE for the two-step method. This again suggests the two-step method may carry more biases when the sample size increases and might not be a valid method to follow. The estimation performance is slightly better in the lower dimensional scenario $(p,q)=(10,10)$ compared to the high-dimensional setting where $(p,q)=(10,50)$ for both joint model and two-step method. This indicates it is easier to capture the true mediation effects when the dimension of the matrix-valued data is lower, but both methods can still handle the high-dimensional case and perform fairly well. We note that most of the biases for the joint model are negative, and in the small sample size $n=100$ setting, they are slightly larger than the two-step method. This is because the weakly informative priors penalize large estimated parameters, causing the joint model to slightly underestimate causal decomposition effects. These biases are reduced when increasing sample size to $n=300$, suggesting convergence, and making them smaller than the two-step method. Non-informative priors also significantly reduce the biases of the joint model (see additional results in Web Appendix \ref{appdx:comparativeStudy_results}, Web Table \ref{tb:comparativeStudy}), making the biases lower than the two-step method even in the $n=100$ setting. We also note that as the sample size increases, the difference in biases, MSEs and variances between the joint model and the two-step method decreases for all decomposition effects. Overall, the joint model improves the efficiency of estimating NIE, NDE, and TE compared to the two-step method, especially in small sample size settings. 

\begin{figure}[ht!]
    \centering
    \includegraphics[width=\textwidth]{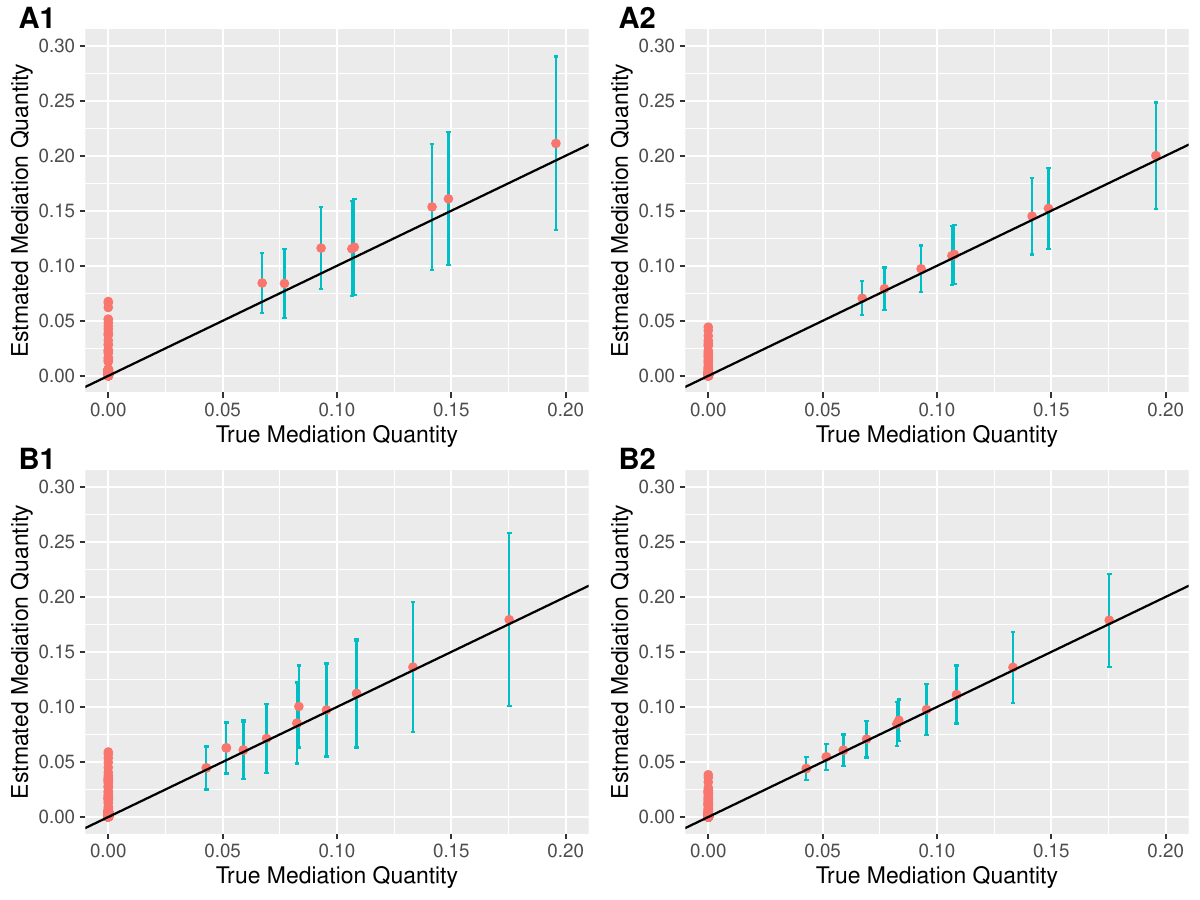}
        \caption{Scatter plots of the mean estimated mediation quantities (red dots) with the black diagonal line and blue error bars (estimates $\pm$ 1-standard deviation across 500 simulation replicates) in the simulation. Panel A1: $(p,q)=(10,10)$, $n=100$; B1: $(p,q)=(10,10)$, $n=300$; A2: $(p,q)=(10,50)$, $n=100$; B2: $(p,q)=(10,50)$, $n=300$.}
    \label{fig:Scatter}
\end{figure}

Figure \ref{fig:Scatter} shows the mean estimated mediation quantities \eqref{eq:med} over 500 simulation replicates. In the low dimensional setting $(p,q)=(10,10)$, for both small (panel A1) and large (panel A2) sample sizes, the joint model can completely differentiate between active indicators (true quantity $>$ 0) and inactive indicators (true quantity = 0), which gives AUC values 1 for both scenarios. In the high-dimensional setting $(p,q)=(10,50)$, although a few active indicators mixed up with inactive indicators in the small sample size setting (Panel B1), most active indicators are still distinguishable, giving an AUC of 0.99. In the large sample size setting (Panel B2), the model also gives AUC 1, meaning it can completely differentiate active indicators. As the sample size $n$ increases from 100 to 300, more estimates for inactive indicators approach 0, and the gap of estimates between active and inactive indicators becomes more significant, indicating better differentiation. Increasing sample size also reduces estimation errors for active indicators, as seen by the smaller error bars. Moreover, the estimates are closer to the true values, as seen by the red dots approaching the black diagonal line, and the order of these estimates also approaches the correct order of true mediation quantities.

\begin{figure}[ht!]
    \centering
    \includegraphics[width=\textwidth]{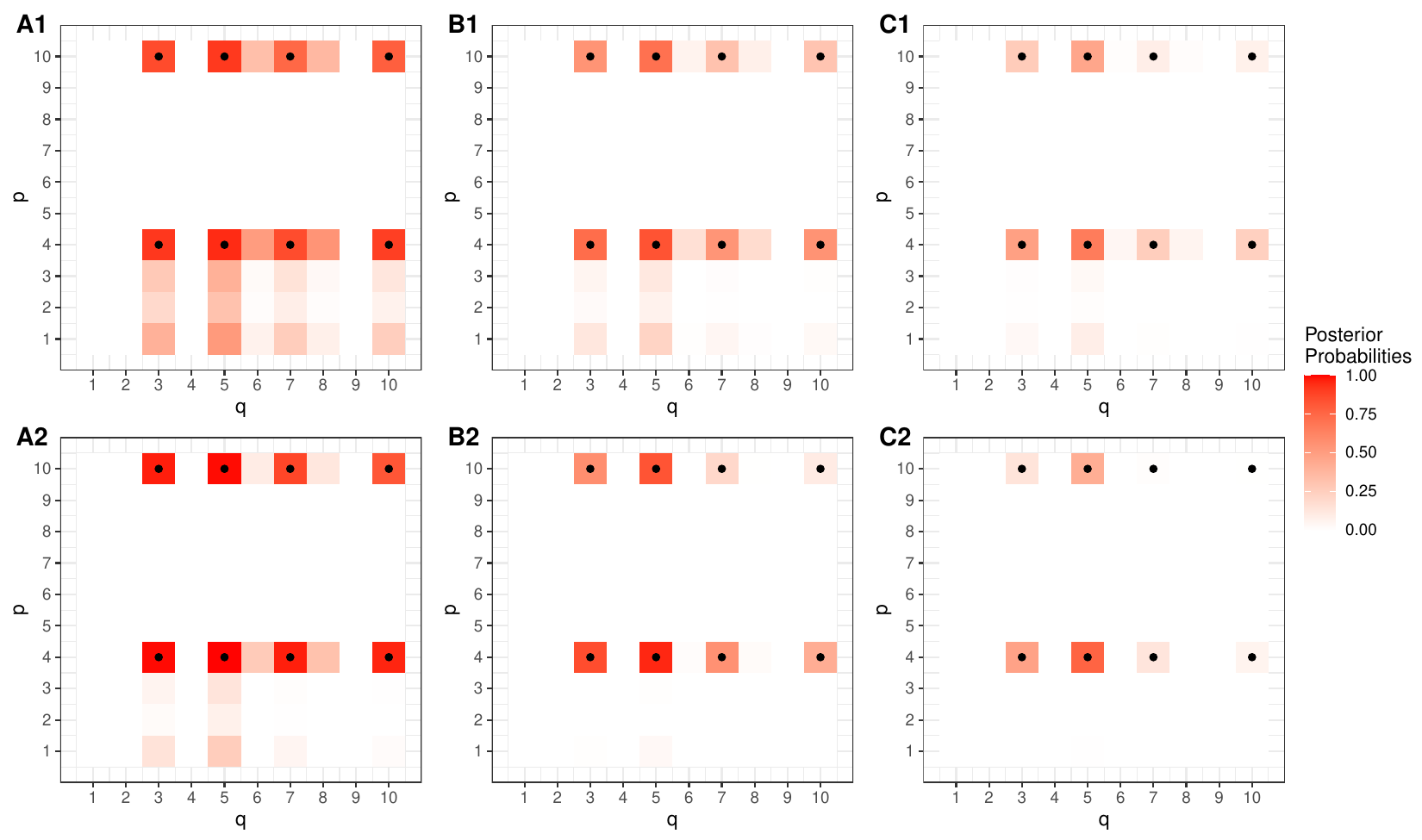}
        \caption{Heatmaps of posterior probabilities in the simulation. A larger value indicates a higher probability of mediation. Black dots: Simulated active indicators of mediation. Panels A1, B1, C1: Thresholds $\kappa=$ 0.05, 0.10, 0.15, $n=100$; Panels A2, B2, C2: Thresholds $\kappa=$ 0.05, 0.10, 0.15, $n=300$.}
    \label{fig:BayesianP_p10q10}
\end{figure}

\begin{figure}[ht!]
    \centering
    \includegraphics[width=\textwidth]{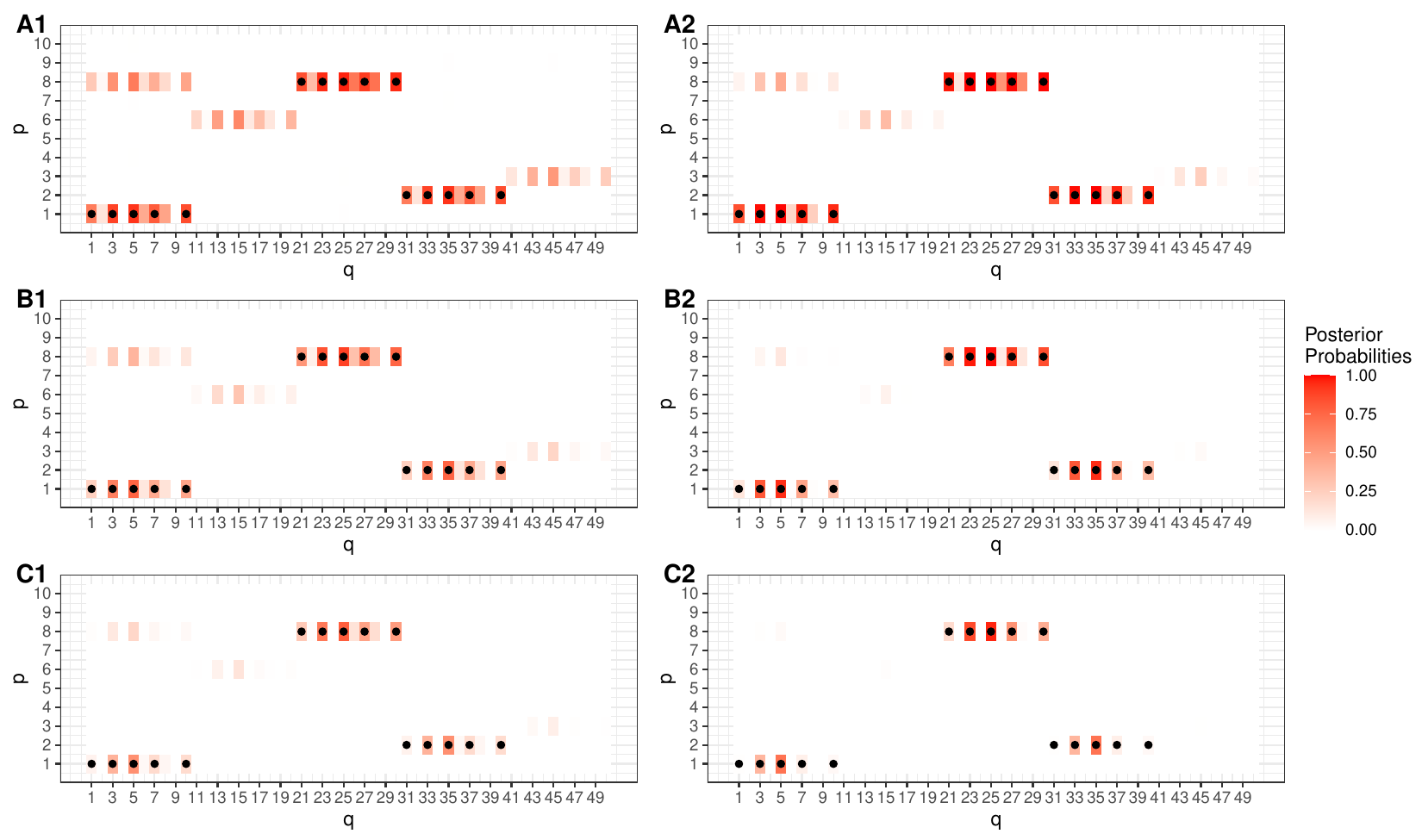}
        \caption{Heatmaps of posterior probabilities in the simulation. A larger value indicates a higher probability of mediation. Black dots: Simulated active indicators of mediation. Panels A1, B1, C1: Thresholds $\kappa=$ 0.03, 0.06, 0.09, $n=100$; Panels A2, B2, C2: Thresholds $\kappa=$ 0.03, 0.06, 0.09, $n=300$.}
    \label{fig:BayesianP_p10q50}
\end{figure}

Figures \ref{fig:BayesianP_p10q10} and \ref{fig:BayesianP_p10q50} show the mean posterior probabilities across 500 simulation replicates, with panel labels A1, B1 and C1 indicating small, medium, and large $\kappa=$ thresholds for $n=100$, and A2, B2 and C2 indicating those three levels of thresholds for $n=300$, respectively. For both low and high-dimensional settings, in the small $\kappa$ and $n=100$ setting (Panel A1), all active indicators can be identified with large posterior probabilities. However, posterior probabilities for certain inactive indicators are also relatively large, and these inactive indicators show similar block patterns as the active indicators, especially in the high-dimensional setting. Increasing the sample size to 300 (Panel A2) can reduce this noise somewhat. In the medium $\kappa$ and $n=100$ setting (Panel B1), the noise for inactive indicators reduces as well, and active indicators are still distinguishable. Increasing the sample size to 300 (Panel B2) further reduces false positive indicators while keeping the active indicators identifiable. When $\kappa$ is large and $n=100$ (Panel C1), all posterior probabilities decrease compared to settings $n=100$ with smaller thresholds, resulting in reduced noise but also less clear results for the active indicators. Increasing $n$ to 300 (Panel C2) almost eliminates all false positive indicators, but some active indicators become indistinguishable. Overall, increasing the threshold can diminish identifying false positive indicators but also lower the chances of identifying active indicators. Increasing the sample size helps in reducing noise and increasing the gap of posterior probabilities between active and inactive indicators. These results are similar under both low dimensional ($p=10,q=50$) and high-dimensional ($p=10,q=50$) scenarios.

\section{Application}
We apply the proposed methods in data collected in a prospective cohort study. A total of 101 patients with anal or perinatal cancer were enrolled between 2008 and 2013, and were treated with concurrent chemotherapy and image-guided intensity-modulated radiation therapy with prescription dose ranging from 45 to 63 Gy to gross targets \citep{lukovic2023evaluation}. Among these, $n=87$ patients were evaluated for this mediation analysis, with 28 patients (32.2\%) experiencing at least one unplanned treatment interruption ($Y_i=1$), and 59 patients (67.8\%) did not experience any treatment interruption ($Y_i=0$). Of the 87 patients, 50 (57.5\%) received a prescription dose $>$ 54 Gy ($E_i=1$), while 37 patients (42.5\%) received a prescription dose $\leq$ 54 Gy ($E_i=0$). DVH data were extracted from a clinical treatment planning system. Dosimetric parameters V1 to V69 were extracted in 1 Gy increment ($q=69$) from DVH (where V$x$ denotes the percentage volume receiving at least $x$ Gy) for each of the 7 OARs: bladder, skin, genital organ (vagina or penile bulb), rectum, large bowel, small bowel, and sphincter ($p=7$). 

The matrix-valued DVH data $\bm{X}_i$ were first centered by the sample mean matrix $\overline{\bm{X}} = n^{-1} \sum_{i=1}^{n}\bm{X}_i$. Then, each element of $\bm{X}_i$ (denoted by $X_{i, j k}$, where $i=1,\cdots,n$, $j=1,\cdots,p$, $k=1,\cdots,q$) was scaled by the element-wise standard deviation ($\sqrt{n^{-1} \sum_{i=1}^{n} (X_{i, j k} - \overline{X}_{j k})^2}$, where $\overline{X}_{j k} = n^{-1} \sum_{i=1}^{n} X_{i, j k} $). Scaling helps our joint model to capture small variations in some OARs and doses, and also reduces the influence of those OARs and doses with very large variations relative to others. The choice of $p_0$ and $q_0$ in the joint model is described in Web Appendix \ref{appdx:appStudy_DIC}. We considered a total of 18 models, specified by combinations of $(p_0,q_0)$ where $p_0 \in \{4,5,6\}$ and $q_0 \in \{4,5,6,7,8,9\}$. For each model, we ran 25,000 iterations of MCMC and retained every 5th sample, with an extra 7500 iterations of burn-in period, resulting in 5000 posterior samples in total. 

\begin{figure}
    \centering
    \includegraphics[width=\textwidth]{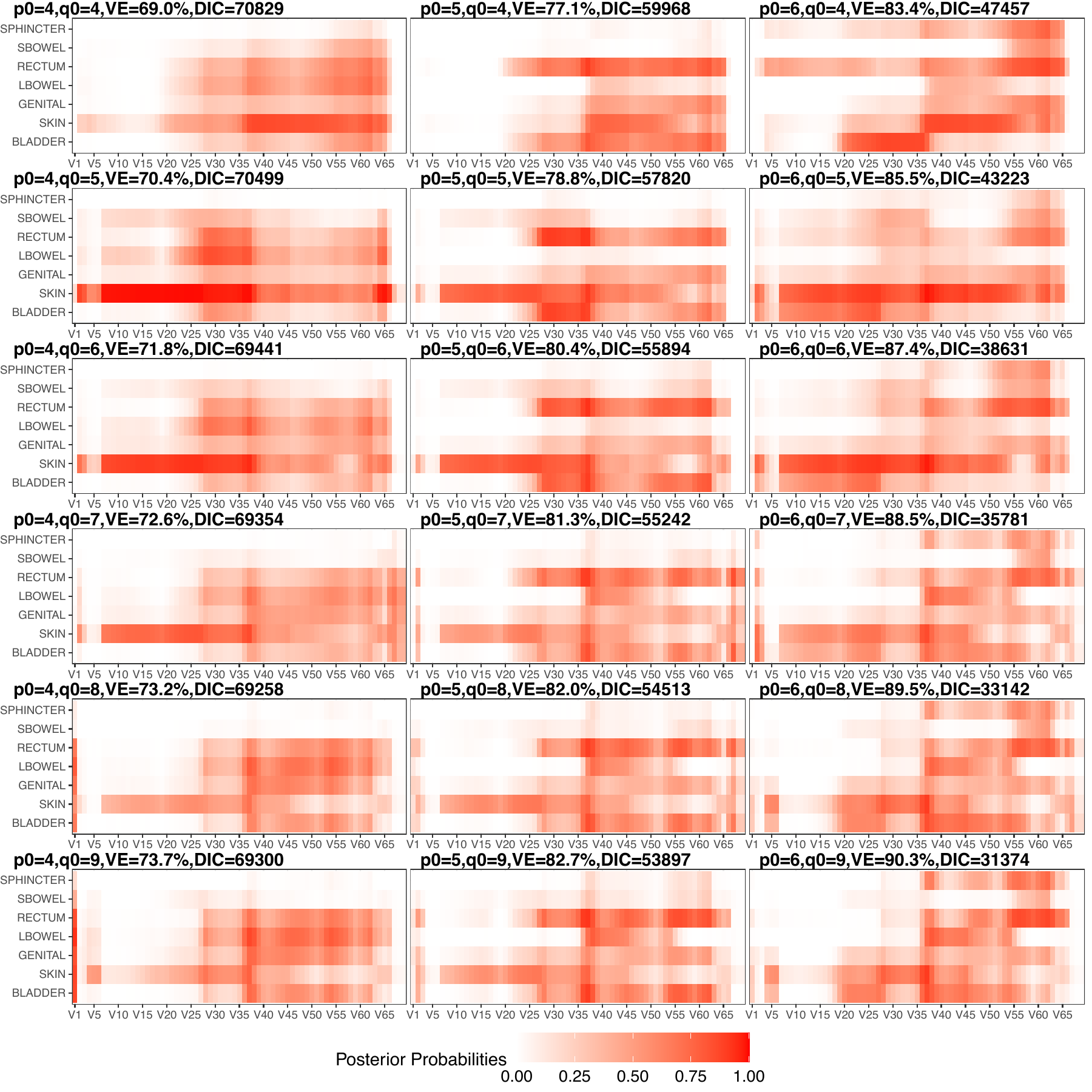}
        \caption{Heatmaps of posterior probabilities for several choices of $p_0$ and $q_0$ in the DVH application. Threshold $\kappa = 75\%$ percentile of the range of the posterior mean mediation quantities for each scenario. VE = variance explained by latent features, SBOWEL = small bowel, LBOWEL = large bowel.}
    \label{fig:DVH_BayesianP_75kappa}
\end{figure}

Figure \ref{fig:DVH_BayesianP_75kappa} presents heatmaps of posterior probabilities, the percentages of variance that can be explained by latent features (VE), and DIC for all the considered models. We observe that increasing either $p_0$ or $q_0$ leads to higher variance explained by latent features. More than $69\%$ of the variance could be explained by the MPCA components across all the models considered, with the highest percentage exceeding 90\% ($p_0=6, q_0=9$). A relatively small number of features ($p_0=5, q_0=6$) still achieves an explained variance $> 80\%$. For the models considered, DIC keeps decreasing when increasing $p_0$ and $q_0$, favoring the largest model with $p_0=6, q_0=9$. 

A higher posterior probability (deeper red in Figure \ref{fig:DVH_BayesianP_75kappa}) indicates a higher probability of an OAR within a specific dose range mediating the causal relationship from the prescription dose to treatment interruption. For each model, the threshold $\kappa$ is selected to be 75\% percentile of the posterior means of the estimated mediation quantities. Heatmaps for $\kappa=50\%$ and $\kappa=25\%$ can be found in Web Appendix \ref{appdx:appStudy_figures}. As we can see from Figure \ref{fig:DVH_BayesianP_75kappa}, increasing $p_0$ (the number of row-wise features for OARs) enhances the separation of posterior probabilities between different OARs. For example, in the model with $p_0=4, q_0=4$, the posterior probability patterns do not vary significantly between different rows (OARs). However, when we increase $p_0$ to 6 while keeping $q_0$ at 4, a clearer separation between OARs is observed, as sphincter lights up and small and large bowel disappear at lower doses. Similarly, increasing $q_0$ (the number of column-wise features for doses) captures more variation from different doses, as the patterns have changed for middle doses and higher doses from $p_0=4, q_0=4$ to $p_0=4, q_0=9$. Across almost all models, the middle doses (25-40 Gy) to skin and larger doses (50-65 Gy) to rectum appear to have higher mediation probabilities.

 Because Figure \ref{fig:DVH_BayesianP_75kappa} showed that beyond $(p_0, q_0) = (6,8)$ there is relatively little change in the heatmap patterns, VE and DIC, we chose $(p_0, q_0) = (6,8)$ as as our final model.  Based on this model, dose ranges 55-65 Gy to rectum, 35-45 Gy to large bowel, 28-38 Gy to skin, 20-28 Gy and 35-58 Gy to bladder are associatied with mediation. The posterior means of causal decomposition effects and their 90\% credible intervals (CI) in this scenario are NIE = 0.142 (90\% CI: 0.028, 0.257), NDE = -0.010 (90\% CI: -0.092, 0.075), and TE = 0.132 (90\% CI: 0.015, 0.247). This indicates that DVH significantly mediates the causal relationship from the prescription dose to the outcome overall, since NIE is significant, while no significant direct effect was observed in the analysis. 

\section{Discussion}
\label{sec:Discussion}

In this paper, we proposed a novel three-component joint model for mediation analysis with high-dimensional, highly correlated, matrix-valued data. We formulated the causal decomposition effects through pathways involving latent factors, proposed a simple Varimax rotation procedure to identify the active indicators of mediation, and adopted a Gibbs sampling algorithm to simultaneously estimate all model parameters. Our simulation study demonstrated that the proposed joint model has higher efficiency estimating causal decomposition effects compared to an alternative two-step method, and demonstrated the mediation effects can be identified and visualized in the original matrix form. 

We summarize several key contributions of our method. (i) Our joint model was developed specifically for mediation analysis with matrix-valued data, distinct from other high-dimensional mediation analysis methods. Although the MPCA-based feature extraction is similar to how other literature handled matrix-valued data \citep{jiang2020bayesian, ding2014dimension}, we incorporated MPCA as a component in a joint model specifically for the purpose of mediation analysis. (ii) To identify active indicators of mediation, we proposed a simple Varimax rotation method, which is computationally more straightforward than the L1-penalization used in \citet{derkach2019high}, while still giving identifiable results when the model is correctly specified. (iii) We used a Bayesian estimation method, which allows quantifying uncertainties in all estimated model parameters and effects. These uncertainties are further summarized by posterior probabilities visualized in the original matrix form, offering a probabilistic interpretation for the matrix-valued mediation effects.

We emphasize that the proposed joint model is a latent variable mediation model, which is the reason we referred to the elements of the matrix-valued data as \say{indicators of mediation} rather than \say{mediators}. However, we note that the joint model (\ref{equ:JointMedMPCA}-\ref{equ:JointMedProbit}) implies the following models for the observed data: (i) a linear model regressing the observed matrix-valued data $\bm{X}_i$ on $E_i$ and $\bm{Z}_i$; (ii) a probit model regressing $Y_i$ on $\bm{X}_i$, $E_i$ and $\bm{Z}_i$. These two implied models reflect the association structure between $Y_i$, $E_i$, $\bm{Z}_i$ and $\bm{X}_i$, marginal over $\bm{T}_i$. However, because the causal effects in the joint model were defined in terms of the latent features $\bm{T}_i$, these associations do not generally identify causal effects. The implied models are derived, and their interpretation further discussed in Web Appendix \ref{appdx:impliedModel}.

In our real data analysis, the total effect of the RT target prescription dose on the outcome can be broken into two parts: an indirect effect of the prescription dose on the outcome that works through the mediator of OAR planned doses, and a direct effect on the outcome through pathways other than this mediator. The direct effect can involve pathways through other (possibly unobserved) mediators, for example the delivered OAR doses, which could be different from the OAR planned doses. The analysis identified significant indirect effects, but did not detect significant direct effect, suggesting that the treatment effects are mediated through certain OARs and dose ranges. The middle doses of skin are identified as strong indicators of mediation in several scenarios with different number of features, which is consistent with previous research where skin toxicity was found to be significantly correlated with certain dosimetric parameters and treatment breaks \citep{lukovic2023evaluation}. The analysis suggests that the number of latent features in the joint model should be chosen large enough so the model can capture variation relevant to mediation. The joint model can accommodate a larger number of latent features because of the mildly informative priors, while the two-step method might encounter convergence problems if incorporating too many features into the regression model. 

Our joint model can be applied to any other matrix-valued data beyond DVH data, or even higher-mode tensor data, by replacing $\bm{B} \otimes \bm{A}$ in the vectorized formulation of equation \eqref{equ:JointMedMPCA} with one projection matrix for each mode. We distinguish our work from other recent proposal that either vectorize the matrix mediator and penalize regression coefficients using a likelihood-based approach \citep{chen2023causal}, or can only be applied to a more constrained matrix mediator in some specialized application areas, such as a symmetrical matrix mediator \citep{zhao2022bayesian} or a covariance matrix mediator \citep{xu2023mediation} for brain functional connectomes, or a mammogram image matrix mediator with irregular boundaries \citep{jiang2023causal}. In general, our joint model is suitable for general matrix mediators, where each subject's columns and rows contain the same information (e.g., image matrices need to be well-aligned).

We note that the proposed Gibbs sampling procedure for the joint model can only handle the case when both row and column dimensions of the matrix-valued data are reduced, i.e. $p_0<p$ and $q_0<q$. This is because the loading matrices $\bm{A}$ and $\bm{B}$ are updated column-by-column, and the full conditional posterior distribution of each column will remain fixed up to sign changes if $p=p_0$ or $q=q_0$, causing the Markov chain to be reducible \citep{hoff2009simulation}. If one dimension of the matrix-valued data is reduced, MPCA is still a different model from PPCA because MPCA reduces the number of parameters by ($pq \times p_0 q_0) - (pp_0 + qq_0)$, achieved through the Kronecker envelope. We acknowledge that $p_0$ or $q_0$ might be as large as $p$ or $q$ for certain applications. For example, the dimension of OAR in the DVH application might not need to be reduced if only a few OARs are present. While our proposed updating algorithm requires choosing $p_0 < p$ and $q_0 < q$, in future work, we are seeking an alternative algorithm, based on Hamiltonian MC instead of Gibbs sample, that can jointly update the loading matrices to remove the need for this restriction. As another potential limitation, we assumed that the model is correctly specified. If the true loading matrix does not correspond to the Varimax solution, bias may result in estimating the quantities \eqref{eq:med}. However, additional simulations (results not shown) suggested that the proposed method is still useful for detecting active indicators of mediation as long as the true loading matrix is sparse. Alternative methods could involve considering other orthogonal rotation methods, or placing shrinkage priors on the loadings similar to \citet{song2020bayesian,song2021bayesian} to obtain sparseness. Finally, although the model was here formulated for a binary outcome, it can be easily fitted to a continuous outcome by just removing the probit link. Generalizing to other outcomes like ordinal and survival outcomes needs further work on conjugate priors to ensure efficient Gibbs sampling, or different sampling algorithms.

\section*{Acknowledgements}

The work of the first author was supported by Artificial Intelligence for Public Health (AI4PH) Health Research Training Platform (HRTP) and the work of the last author by a Discovery Grant from the Natural Sciences and Engineering Research Council of Canada.


\section*{Supplementary Materials}

Web Appendices, Tables, and Figures referenced in Section \ref{sec:intro}-\ref{sec:Discussion}, are available on the same arXiv web page with this manuscript in the Ancillary files section.

\section*{Data Availability}

The data used for demonstrating the methods were housed and analyzed at Princess Margaret Cancer Centre, University Health Network. They cannot be publicly shared due to confidentiality/privacy reasons.

\bibliographystyle{apalike}
\bibliography{bibliography_joint_model}

\begin{thebibliography}{}

\bibitem[Bauer et~al., 2006]{bauer2006principal}
Bauer, J., Jackson, A., Skwarchuk, M., and Zelefsky, M. (2006).
\newblock Principal component, {V}arimax rotation and cost analysis of volume
  effects in rectal bleeding in patients treated with {3D-CRT} for prostate
  cancer.
\newblock {\em Physics in Medicine \& Biology}, 51(20):5105.

\bibitem[Chen and Zhou, 2023]{chen2023causal}
Chen, M. and Zhou, Y. (2023).
\newblock Causal mediation analysis with a three-dimensional image mediator.
\newblock {\em Statistics in Medicine}.

\bibitem[Ch{\'e}n et~al., 2018]{chen2018high}
Ch{\'e}n, O.~Y., Crainiceanu, C., Ogburn, E.~L., Caffo, B.~S., Wager, T.~D.,
  and Lindquist, M.~A. (2018).
\newblock High-dimensional multivariate mediation with application to
  neuroimaging data.
\newblock {\em Biostatistics}, 19(2):121--136.

\bibitem[Clark-Boucher et~al., 2023]{clark2023methods}
Clark-Boucher, D., Zhou, X., Du, J., Liu, Y., Needham, B.~L., Smith, J.~A., and
  Mukherjee, B. (2023).
\newblock Methods for mediation analysis with high-dimensional dna methylation
  data: Possible choices and comparisons.
\newblock {\em PLoS genetics}, 19(11):e1011022.

\bibitem[Dawson et~al., 2005]{dawson2005use}
Dawson, L.~A., Biersack, M., Lockwood, G., Eisbruch, A., Lawrence, T.~S., and
  Ten~Haken, R.~K. (2005).
\newblock Use of principal component analysis to evaluate the partial organ
  tolerance of normal tissues to radiation.
\newblock {\em International Journal of Radiation Oncology, Biology, Physics},
  62(3):829--837.

\bibitem[Dean et~al., 2016]{dean2016functional}
Dean, J.~A., Wong, K.~H., Gay, H., Welsh, L.~C., Jones, A.-B., Schick, U., Oh,
  J.~H., Apte, A., Newbold, K.~L., Bhide, S.~A., et~al. (2016).
\newblock Functional data analysis applied to modeling of severe acute
  mucositis and dysphagia resulting from head and neck radiation therapy.
\newblock {\em International Journal of Radiation Oncology, Biology, Physics},
  96(4):820--831.

\bibitem[Derkach et~al., 2019]{derkach2019high}
Derkach, A., Pfeiffer, R.~M., Chen, T.-H., and Sampson, J.~N. (2019).
\newblock High dimensional mediation analysis with latent variables.
\newblock {\em Biometrics}, 75(3):745--756.

\bibitem[Ding and Cook, 2014]{ding2014dimension}
Ding, S. and Cook, R.~D. (2014).
\newblock Dimension folding {PCA} and {PFC} for matrix-valued predictors.
\newblock {\em Statistica Sinica}, 24(1):463--492.

\bibitem[Ding and Cook, 2018]{ding2018matrix}
Ding, S. and Cook, R.~D. (2018).
\newblock Matrix variate regressions and envelope models.
\newblock {\em Journal of the Royal Statistical Society Series B: Statistical
  Methodology}, 80(2):387--408.

\bibitem[Gao et~al., 2019]{gao2019testing}
Gao, Y., Yang, H., Fang, R., Zhang, Y., Goode, E.~L., and Cui, Y. (2019).
\newblock Testing mediation effects in high-dimensional epigenetic studies.
\newblock {\em Frontiers in genetics}, 10:1195.

\bibitem[Hendry et~al., 1996]{hendry1996modelled}
Hendry, J.~H., Bentzen, S.~M., Dale, R., Fowler, J.~F., Wheldon, T., Jones, B.,
  Munro, A., Slevin, N.~J., and Robertson, A.~G. (1996).
\newblock A modelled comparison of the effects of using different ways to
  compensate for missed treatment days in radiotherapy.
\newblock {\em Clinical Oncology}, 8(5):297--307.

\bibitem[Hoff, 2007]{hoff2007model}
Hoff, P.~D. (2007).
\newblock Model averaging and dimension selection for the singular value
  decomposition.
\newblock {\em Journal of the American Statistical Association},
  102(478):674--685.

\bibitem[Hoff, 2009]{hoff2009simulation}
Hoff, P.~D. (2009).
\newblock Simulation of the matrix {B}ingham--von {M}ises--{F}isher
  distribution, with applications to multivariate and relational data.
\newblock {\em Journal of Computational and Graphical Statistics},
  18(2):438--456.

\bibitem[Hoff, 2015]{hoff2015multilinear}
Hoff, P.~D. (2015).
\newblock Multilinear tensor regression for longitudinal relational data.
\newblock {\em The annals of applied statistics}, 9(3):1169.

\bibitem[Huang and Pan, 2016]{huang2016hypothesis}
Huang, Y.-T. and Pan, W.-C. (2016).
\newblock Hypothesis test of mediation effect in causal mediation model with
  high-dimensional continuous mediators.
\newblock {\em Biometrics}, 72(2):402--413.

\bibitem[Hung et~al., 2012]{hung2012multilinear}
Hung, H., Wu, P., Tu, I., and Huang, S. (2012).
\newblock On multilinear principal component analysis of order-two tensors.
\newblock {\em Biometrika}, 99(3):569--583.

\bibitem[Jiang et~al., 2020]{jiang2020bayesian}
Jiang, B., Petkova, E., Tarpey, T., and Ogden, R.~T. (2020).
\newblock A {B}ayesian approach to joint modeling of matrix-valued imaging data
  and treatment outcome with applications to depression studies.
\newblock {\em Biometrics}, 76(1):87--97.

\bibitem[Jiang and Colditz, 2023]{jiang2023causal}
Jiang, S. and Colditz, G.~A. (2023).
\newblock Causal mediation analysis using high-dimensional image mediator
  bounded in irregular domain with an application to breast cancer.
\newblock {\em Biometrics}, 79(4):3728--3738.

\bibitem[Li and Zhang, 2017]{li2017parsimonious}
Li, L. and Zhang, X. (2017).
\newblock Parsimonious tensor response regression.
\newblock {\em Journal of the American Statistical Association},
  112(519):1131--1146.

\bibitem[Lu et~al., 2008]{lu2008mpca}
Lu, H., Plataniotis, K.~N., and Venetsanopoulos, A.~N. (2008).
\newblock {MPCA}: Multilinear principal component analysis of tensor objects.
\newblock {\em IEEE transactions on Neural Networks}, 19(1):18--39.

\bibitem[Lukovic et~al., 2023]{lukovic2023evaluation}
Lukovic, J., Hosni, A., Liu, A., Chen, J., Tadic, T., Patel, T., Li, K., Han,
  K., Lindsay, P., Craig, T., et~al. (2023).
\newblock Evaluation of dosimetric predictors of toxicity after {IMRT} with
  concurrent chemotherapy for anal cancer.
\newblock {\em Radiotherapy and Oncology}, 178:109429.

\bibitem[O'Shea et~al., 2022]{o2022compensation}
O'Shea, K., Coleman, L., Fahy, L., Kleefeld, C., Foley, M.~J., and Moore, M.
  (2022).
\newblock Compensation for radiotherapy treatment interruptions due to a
  cyberattack: An isoeffective {DVH}-based dose compensation decision tool.
\newblock {\em Journal of Applied Clinical Medical Physics}, 23(9):e13716.

\bibitem[Rohe and Zeng, 2023]{rohe2020vintage}
Rohe, K. and Zeng, M. (2023).
\newblock Vintage factor analysis with varimax performs statistical inference.
\newblock {\em Statistical Society Series B: Statistical Methodology},
  85(4):1037--1060.

\bibitem[Skala et~al., 2007]{skala2007patient}
Skala, M., Rosewall, T., Dawson, L., Divanbeigi, L., Lockwood, G., Thomas, C.,
  Crook, J., Chung, P., Warde, P., and Catton, C. (2007).
\newblock Patient-assessed late toxicity rates and principal component analysis
  after image-guided radiation therapy for prostate cancer.
\newblock {\em International Journal of Radiation Oncology, Biology, Physics},
  68(3):690--698.

\bibitem[Song et~al., 2021]{song2021bayesian}
Song, Y., Zhou, X., Kang, J., Aung, M.~T., Zhang, M., Zhao, W., Needham, B.~L.,
  Kardia, S.~L., Liu, Y., Meeker, J.~D., et~al. (2021).
\newblock Bayesian sparse mediation analysis with targeted penalization of
  natural indirect effects.
\newblock {\em Journal of the Royal Statistical Society. Series C, Applied
  statistics}, 70(5):1391.

\bibitem[Song et~al., 2020]{song2020bayesian}
Song, Y., Zhou, X., Zhang, M., Zhao, W., Liu, Y., Kardia, S.~L., Roux, A.
  V.~D., Needham, B.~L., Smith, J.~A., and Mukherjee, B. (2020).
\newblock Bayesian shrinkage estimation of high dimensional causal mediation
  effects in omics studies.
\newblock {\em Biometrics}, 76(3):700--710.

\bibitem[Tipping and Bishop, 1999]{tipping1999probabilistic}
Tipping, M.~E. and Bishop, C.~M. (1999).
\newblock Probabilistic principal component analysis.
\newblock {\em Journal of the Royal Statistical Society: Series B (Statistical
  Methodology)}, 61(3):611--622.

\bibitem[Xu and Zhao, 2023]{xu2023mediation}
Xu, Y. and Zhao, Y. (2023).
\newblock Mediation analysis with graph mediator.
\newblock {\em arXiv preprint arXiv:2307.03821}.

\bibitem[Zhang et~al., 2016]{zhang2016estimating}
Zhang, H., Zheng, Y., Zhang, Z., Gao, T., Joyce, B., Yoon, G., Zhang, W.,
  Schwartz, J., Just, A., Colicino, E., et~al. (2016).
\newblock Estimating and testing high-dimensional mediation effects in
  epigenetic studies.
\newblock {\em Bioinformatics}, 32(20):3150--3154.

\bibitem[Zhang, 2021]{zhang2021high}
Zhang, Q. (2021).
\newblock High-dimensional mediation analysis with applications to causal gene
  identification.
\newblock {\em Statistics in Biosciences}, pages 1--20.

\bibitem[Zhao and Leng, 2014]{zhao2014structured}
Zhao, J. and Leng, C. (2014).
\newblock Structured lasso for regression with matrix covariates.
\newblock {\em Statistica Sinica}, pages 799--814.

\bibitem[Zhao et~al., 2022]{zhao2022bayesian}
Zhao, Y., Chen, T., Cai, J., Lichenstein, S., Potenza, M.~N., and Yip, S.~W.
  (2022).
\newblock Bayesian network mediation analysis with application to the brain
  functional connectome.
\newblock {\em Statistics in medicine}, 41(20):3991--4005.

\bibitem[Zhao et~al., 2020]{zhao2020sparse}
Zhao, Y., Lindquist, M.~A., and Caffo, B.~S. (2020).
\newblock Sparse principal component based high-dimensional mediation analysis.
\newblock {\em Computational statistics \& data analysis}, 142:106835.

\bibitem[Zhou and Li, 2014]{zhou2014regularized}
Zhou, H. and Li, L. (2014).
\newblock Regularized matrix regression.
\newblock {\em Journal of the Royal Statistical Society. Series B, Statistical
  Methodology}, 76(2):463.

\bibitem[Zhou et~al., 2013]{zhou2013tensor}
Zhou, H., Li, L., and Zhu, H. (2013).
\newblock Tensor regression with applications in neuroimaging data analysis.
\newblock {\em Journal of the American Statistical Association},
  108(502):540--552.

\end{thebibliography}


\begin{thebibliography}{}

\bibitem[Anderson and Rubin, 1956]{anderson1956statistical}
Anderson, T. and Rubin, H. (1956).
\newblock Statistical inference in.
\newblock In {\em Proceedings of the Berkeley Symposium on Mathematical
  Statistics and Probability}, page 111. University of California Press.

\bibitem[Celeux et~al., 2006]{celeux2006deviance}
Celeux, G., Forbes, F., Robert, C.~P., and Titterington, D.~M. (2006).
\newblock Deviance information criteria for missing data models.
\newblock {\em Bayesian Analysis}, 1(4):651 -- 673.

\bibitem[Derkach et~al., 2019]{derkach2019high}
Derkach, A., Pfeiffer, R.~M., Chen, T.-H., and Sampson, J.~N. (2019).
\newblock High dimensional mediation analysis with latent variables.
\newblock {\em Biometrics}, 75(3):745--756.

\bibitem[Ding and Cook, 2014]{ding2014dimension}
Ding, S. and Cook, R.~D. (2014).
\newblock Dimension folding {PCA} and {PFC} for matrix-valued predictors.
\newblock {\em Statistica Sinica}, 24(1):463--492.

\bibitem[Hoff, 2007]{hoff2007model}
Hoff, P.~D. (2007).
\newblock Model averaging and dimension selection for the singular value
  decomposition.
\newblock {\em Journal of the American Statistical Association},
  102(478):674--685.

\bibitem[Hung et~al., 2012]{hung2012multilinear}
Hung, H., Wu, P., Tu, I., and Huang, S. (2012).
\newblock On multilinear principal component analysis of order-two tensors.
\newblock {\em Biometrika}, 99(3):569--583.

\bibitem[Jiang et~al., 2020]{jiang2020bayesian}
Jiang, B., Petkova, E., Tarpey, T., and Ogden, R.~T. (2020).
\newblock A {B}ayesian approach to joint modeling of matrix-valued imaging data
  and treatment outcome with applications to depression studies.
\newblock {\em Biometrics}, 76(1):87--97.

\bibitem[Lange et~al., 2014]{lange2014assessing}
Lange, T., Rasmussen, M., and Thygesen, L.~C. (2014).
\newblock Assessing natural direct and indirect effects through multiple
  pathways.
\newblock {\em American journal of epidemiology}, 179(4):513--518.

\bibitem[Li et~al., 2010]{li2010dimension}
Li, B., Kim, M.~K., and Altman, N. (2010).
\newblock {On dimension folding of matrix- or array-valued statistical
  objects}.
\newblock {\em The Annals of Statistics}, 38(2):1094 -- 1121.

\bibitem[Lu et~al., 2008]{lu2008mpca}
Lu, H., Plataniotis, K.~N., and Venetsanopoulos, A.~N. (2008).
\newblock {MPCA}: Multilinear principal component analysis of tensor objects.
\newblock {\em IEEE transactions on Neural Networks}, 19(1):18--39.

\bibitem[Tipping and Bishop, 1999]{tipping1999probabilistic}
Tipping, M.~E. and Bishop, C.~M. (1999).
\newblock Probabilistic principal component analysis.
\newblock {\em Journal of the Royal Statistical Society: Series B (Statistical
  Methodology)}, 61(3):611--622.

\end{thebibliography}

\end{document}